\documentclass[lettersize,journal]{IEEEtran}
\usepackage[utf8]{inputenc}
\usepackage{graphicx}
\usepackage{multicol}
\usepackage{amsmath}
\usepackage{amsfonts}
\usepackage{amsthm,amssymb}
\usepackage[mathscr]{euscript}
\usepackage{mdwmath}
\usepackage{bbm}
\usepackage{mathrsfs}
\usepackage{bm}
\usepackage{epstopdf}
\usepackage{booktabs}
\usepackage[flushleft]{threeparttable}
\usepackage{siunitx}
\usepackage[tight,footnotesize]{subfigure}
\usepackage{cite}
\usepackage{wrapfig}
\usepackage[justification=centering]{caption}
\usepackage[font={small}]{caption}
\usepackage{algorithm}
\usepackage[]{algpseudocode}
\usepackage{float}
\usepackage{epstopdf}
\usepackage{comment}
\usepackage{marginnote}
\usepackage{color}
\usepackage{caption}
\newcommand{\mydef}{:=}
\newcommand{\Rc}{\mathbb{R}}

\newcommand{\bd}[1]{\mathbf{#1}}
\newcommand{\ngg}{N_g}
\newcommand{\nbb}{N_b}
\newcommand{\ttt}{\theta}

\begin{document}

\title{Approximating Voltage Stability Boundary Under\\ High Variability of Renewables Using \\  Differential Geometry}

\author{Dan Wu,\quad Franz-Erich Wolter,\quad Sijia Geng
\thanks{Dan Wu is with the School of Electrical and Electronic Engineering, Huazhong University of Science and Technology, Wuhan, China.}
\thanks{Franz-Erich Wolter is with the Department of Computer Science,
Leibniz University Hannover, Hannover, Germany.}
\thanks{Sijia Geng is with the Department of Electrical and Computer Engineering,
Johns Hopkins University, Baltimore, MD 21218, USA.}
\thanks{ 
Corresponding Author: Sijia Geng (email: sgeng@jhu.edu).}
}




\maketitle

\begin{abstract}
This paper proposes a novel method rooted in differential geometry to approximate the voltage stability boundary of power systems under high variability of renewable generation. We extract intrinsic geometric information of the power flow solution manifold at a given operating point. Specifically, coefficients of the Levi-Civita connection are constructed to approximate the geodesics of the manifold starting at an operating point along any interested directions that represent possible fluctuations in generation and load. Then, based on the geodesic approximation, we further predict the voltage collapse point by solving a few univariate quadratic equations. 
Conventional methods mostly rely on either expensive numerical continuation at specified directions or numerical optimization.
Instead, the proposed approach constructs the Christoffel symbols of the second kind from the Riemannian metric tensors to characterize the complete local geometry which is then extended to the proximity of the stability boundary with efficient computations. 
As a result, this approach is suitable to handle high-dimensional variability in operating points due to the large-scale integration of renewable resources.
Using various case studies, we demonstrate the advantages of the proposed method and provide additional insights and discussions on voltage stability in renewable-rich power systems.
\end{abstract}


\section{Introduction}
   \label{sec:intro}
Renewable generation plays an important role in achieving carbon neutrality by replacing conventional fossil-fuel-based synchronous generation. However, numerous technical challenges need to be solved to maintain the stability and normal operation of the future renewable-rich power systems \cite{gu2022power}.

This paper focuses on the (long-term) voltage stability problem for power systems that have a very high penetration level of renewable generation. 
Such systems will experience significant variations in operating points that are larger in magnitude, faster in timescale, and higher in dimension\cite{geng2022unified}. 
Traditionally, the (long-term) voltage instability results from load change that exceeds the maximum power transfer limit (i.e., maximum loadability point). At that point, the Jacobian of the power flow equation becomes singular\cite{hiskens2001exploring,canizares2003linear}, and we refer to such points to be on the power flow singular solution space boundary (SSB), or the voltage stability boundary.
Conventional analysis methods aim to find the maximum loadability point by directly computing the parameter that achieves the singularity condition. That is, consider power flow equations $f(x;\lambda)=0$, allow a scale parameter to vary, and find the parameter value when $\det \partial f/\partial x (x;\lambda)=0$. The maximum loadability point further implies the stability margin from the current operating point. 
Another commonly adopted approach is the continuation power flow (CPF) method, which frees up one parameter, for example, representing a single direction of change for the load and generation profile, and seeks the trajectory of solutions of the power flow equation along the direction of varying \cite{iba1991calculation, chiang1995cpflow, chen2003performance, zhang2005continuation, li2008continuation, ghiocel2013power}. Compared to directly computing the maximum loadability point, the CPF method provides the shape of the solution manifold (albeit along the single parameter variation direction). Since each continuation process takes one parameter varying direction, the applicability of the CPF method relies on the predictability of load and generation and a strong engineering understanding of the system in identifying prevalent operation patterns. Such an assumption may no longer hold in renewable-rich power systems that feature numerous possible directions of change within a short period of time \cite{wu2022tri}. To address such a limitation of the CPF method, another line of research focuses on finding the locally closest point (to the nominal parameter) in parameter space that achieves singularity. Optimization techniques were applied in \cite{jung1990marginal} to solve for the smallest margin. Thorough analysis and computation methods were provided in \cite{dobson1993new, alvarado1994computation, dobson1993computing} to compute the locally nearest point in power space. It is worth mentioning that previous research has studied instability due to limit-induced bifurcation, for example, caused by the limit of reactive power supplied by generators \cite{venkatasubramanian1995dynamics,kataoka2005voltage, qiu2018global}. This is out of the scope of the current paper. In this paper, we focus our discussion on {saddle-node bifurcations, i.e., when the Jacobian matrix of the power flow equation encounters singularity} \cite{canizares2003linear, canizares1995conditions}.  
Recent developments have extended the analysis to the Riemannian metric \cite{wu2022tri, wu2021searching} to compute the shortest manifold distance. Although they have the potential to address the challenge of handling numerous power-changing directions, searching for the global stability margin corresponding to all possible power-changing directions is computationally intensive. 
Computational Riemannian geometry has been used as well for treating a closely related problem, i.e., computing local approximations of the SSB manifold \cite{wolter2019differential}. The methods in \cite{wolter2019differential} give numerically precise presentations of the  SSB by geodesic coordinates (i.e., families of geodesics) up to dimension 200 but are computationally costly.

Revealing the global geometry of the power flow manifold and the {\color{black} singularity-induced} voltage stability boundary is beneficial to guiding system operation. In this paper, we propose a completely novel method rooted in mathematical tools from differential geometry. It can replicate the geometry of the power flow manifold in a global manner, that is, along all possible parameter-varying directions, merely from local measurements and efficient computations. 
Specifically, coefficients of the Levi-Civita connection \cite{lee2018:introduction} are constructed to approximate the geodesics of the manifold starting at the operating point along any interested directions that represent possible adjustments in generation and load. The geodesics per se provide valuable information on the shape of the solution manifold. Then, based on the geodesic approximation, we further predict the voltage collapse point by solving the extrema of a few univariate quadratic equations. 
The proposed approach is suitable to handle high-dimensional variability in operating points (potentially due to the large-scale integration of renewable resources), because once the coefficients of the Levi-Civita connection are constructed, the geodesics can be approximated for arbitrary direction.
We demonstrate the advantages of the proposed method and provide additional insights and discussions on voltage stability in renewable-based power systems using various case studies.

\section{Problem Statement}
\label{sec:problem}
\subsection{Power Flow Model}
Consider a connected power grid with in total $\nbb$ nodes. Without loss of generality, assume that the first node is the slack bus, the second to the $\ngg$-th buses are the PV buses (where the active power and voltage magnitude are specified), and the rest nodes are the PQ buses (where the active and reactive power are specified).
 
For the $i$-th PV bus, we have
\begin{eqnarray}
		U^i \sum_{\substack{m=1 \\ m\neq i}}^{\ngg} U^m \big[ G_{im} \cos(\ttt^{im}) + B_{im} \sin(\ttt^{im}) \big] + (U^i)^2 G_{ii} \nonumber \\
		+\, U^i\sum_{\substack{\ngg +1}}^{\nbb} V^n \big[ G_{in} \cos(\ttt^{in}) + B_{in} \sin(\ttt^{in}) \big]
		 = P^i , \label{eq:PV_P}
\end{eqnarray}%
where $U$'s are the voltage magnitudes at PV nodes that are given as constants; $V$'s are the voltage magnitudes at PQ nodes (unknown); $G_{im}$ and $B_{im}$ are parameters of the network, denoting the $(i,m)$-th elements of the bus conductance matrix and bus susceptance matrix, respectively; $\ttt^{im} \mydef \ttt^i - \ttt^m$ is the nodal voltage angle difference and $\ttt^1 = 0$; $P^i$ is the specified active power injections at the $i$-th PV bus. 

Similarly, for the $j$-th PQ bus, we have
\begingroup
\allowdisplaybreaks
\begin{subequations}
	\begin{align}
		& V^j \sum_{m=1}^{\ngg} U^m \big[ G_{jm} \cos(\ttt^{jm}) + B_{jm} \sin(\ttt^{jm}) \big] + (V^j)^2 G_{jj} \nonumber \\
		& + V^j \sum_{\substack{\ngg +1 \\ n \neq j}}^{\nbb} V^n \big[ G_{jn} \cos(\ttt^{jn}) + B_{jn} \sin(\ttt^{jn}) \big] = P^j, \label{eq:PQ_P}\\
		& V^j \sum_{m=1}^{\ngg} U^m \big[ G_{jm} \sin(\ttt^{jm}) - B_{jm} \cos(\ttt^{jm}) \big]  - (V^j)^2 B_{jj} \nonumber \\
		& + V^j \sum_{\substack{\ngg +1 \\ n \neq j}}^{\nbb} V^n \big[ G_{jn} \sin(\ttt^{jn}) - B_{jn} \cos(\ttt^{jn}) \big] = Q^j,  \label{eq:PQ_Q}
	\end{align}\label{eq:PQ}%
\end{subequations}%
\endgroup
where $P^j$ and $Q^j$ are the specified active power and reactive power injections at the $j$-th PQ bus. 

\subsection{Power Flow Manifold}
Equations \eqref{eq:PV_P} and \eqref{eq:PQ} define the power flow map $\mathcal{F}$ that sends nodal voltage magnitudes and angles to the nodal active and reactive power injections. 
\begin{equation}
    \mathcal{F}: \Rc^N \to \Rc^N,~\mathcal{F}(\bd{V}, \bd{\ttt}) = (\bd{P}, \bd{Q}),%
\end{equation}%
where $N = 2 \nbb- \ngg -1$, $\bd{V} \in \Rc^{\nbb -\ngg}$, $\bd{\ttt} \in \Rc^{\nbb-1}$, $\bd{P} \in \Rc^{\nbb-1}$, $\bd{Q} \in \Rc^{\nbb - \ngg}$.

{\color{black} In the traditional power flow problem we assume that the nodal power injections $\bd{P}$ and $\bd{Q}$ are fixed due to the high predictability of conventional generation and load. So, the voltage variables that admit the given power injection comprise a $0$-dimensional point set (the power flow solutions). However, with more renewable generation penetrating into the grid, such an assumption may not be valid in the future. If we relax all the nodal power injections $\bd{P}$ and $\bd{Q}$ as free variables, the point set that follows the power flow map $\mathcal{F}$ yields an $N$-dimensional surface which defines our power flow manifold $\phi$. }
{\color{black}Note that conventional continuation methods only free one parameter up in a single continuation process.}

\subsection{Singularity-Induced Long Term Voltage Stability Boundary}
Long-term voltage instability is strongly related to the power flow Jacobian matrix reaching singularity\cite{ajjarapu1992continuation}. 
Thus, the voltage stability boundary $\partial \phi$ is defined by the point set,
\begin{equation}
    \partial \phi = \{ (\bd{P}, \bd{Q}, \bd{V},\bd{\theta}) | \mathcal{F}(\bd{V}, \bd{\ttt}) = (\bd{P}, \bd{Q}),~\text{det}(\partial \mathcal{F})=0 \}.
\end{equation}%
{\color{black} 
In this paper, we aim to develop a method that can estimate the voltage stability boundary in a global manner, namely, for all possible power variations. We achieve this by constructing the geometric information of the power flow manifold.}

{\color{black}
\subsection{Geometric Interpretation of Long-Term Voltage Stability Region}\label{subsec:interpretation}
By the Whitney Embedding Theorem \cite{mukherjee2015differential}, the $N$-dimensional power flow manifold $\phi$ can be embedded in $\Rc^{2N}$ which is naturally selected as the power-voltage space $\bd{P} \oplus \bd{Q} \oplus \bd{V} \oplus \bd{\ttt}$. The corresponding parameterization map $\bd{r}$ is constructed by stacking the power flow map $\mathcal{F}$ with the identity map $\mathcal{I}$,
\begin{equation}
	\bd{r}:\Rc^{N} \to \Rc^{2N},~\bd{r}(\bd{V}, \bd{\ttt}) = \big(\mathcal{F}, \mathcal{I}\big)(\bd{V}, \bd{\ttt}).%
\end{equation}%
Since the Jacobian matrix of $\bd{r}$ has the full rank of $N$, $\bd{r}$ is regular.

Define a  projection map $\mathcal{P}$ from the power-voltage space to the power subspace,
\begin{equation}
    \mathcal{P}: \bd{P} \oplus \bd{Q} \oplus \bd{V} \oplus \bd{\ttt} \to \bd{P} \oplus \bd{Q}.%
\end{equation}%
If we further restrict $\mathcal{P}$ on $\phi$, denoted as $\mathcal{P}_\phi$, then $\mathcal{P}_\phi: \phi \to \bd{P} \oplus \bd{Q}$ is a map from the $N$-dimensional manifold $\phi$ to $\Rc^N$. 

Let's consider a given operating point $\bd{z}_0 \in \phi$. For any open neighborhood $U$ of $\bd{z}_0$ in $\phi$, if $\mathcal{P}_\phi$ is a homeomorphism on $U$, then 
$(U, \mathcal{P}_\phi)$ composes a so-called \emph{chart} on $\phi$ and the image of $\mathcal{P}_\phi$, denoted as $W$ {\color{black} (in the power space)}, defines a local coordinate system of $U$. 

Note that $\mathcal{P}_\phi$ fails to be a homeomorphism on $U$ if there exists some point $y \in U$ such that the Jacobian matrix of the power flow map $\mathcal{F}$ at $y$ is singular. From differential geometry point of view, this submanifold of singular points defines the boundary of a particular coordinate system $W$ in the power subspace for the largest chart $(U_{\text{max}}, \mathcal{P}_\phi)$. This $U_{\text{max}}$ is exactly the voltage stability region enclosed by the singular boundary. 
}

\section{Metric Tensors and Geodesic Equation}
	\label{sec:geodesic}
In this section, we briefly summarize the differential geometry tools that are used in the paper for self-completeness. Interested readers can refer to \cite{lee2018:introduction,toponogov2006:differential,liseikin2006:computational} for a more comprehensive exposition. Throughout the paper, we adopt the notational convention from differential geometry
unless otherwise clarified. In general, we use superscripts to represent indices and subscripts to represent partial derivatives {\color{black} for multivariable vector-valued functions}. 

\subsection{Tangent Space, Basis and Dual Basis}
Consider a smooth map $\bd{r}: \Rc^n \to \Rc^{n+m}$, 
where $\bd{r}(\bd{s})=\big[r^1(\bd{s}), \dots, r^{n+m}(\bd{s}) \big]$, $\bd{s} = [s^1,\dots,s^n]$, $n > 0$ and $m \ge 0$.
{\color{black} The point set $\phi \mydef \big\{ \bd{r}(\bd{s})\in \Rc^{n+m} ~|~ \bd{s}\in\Rc^n \big\}$ defines an $n$-dimensional surface where $\bd{s}$ is referred to as curvilinear coordinates on $\phi$. The vector-valued function $\bd{r}(\bd{s})$ is a parametric representation for $\phi$. 
}

Define the $i$-th basic tangent vector $\bd{r}_i \in \Rc^{n+m}$ by
\begin{equation}
	\bd{r}_i \mydef \frac{\partial\bd{r}}{\partial s^i}.\label{eq:basic_tangent}
\end{equation}
Let $T\phi_p$ be the tangent space of $\phi$ at some point $p \in \phi$. Then, $T\phi_p$ is spanned by the basis $\{ \bd{r}_i \}$. {\color{black}We say $\phi$ is regular if the rank of the matrix $(\partial r^i/\partial s^j)$ is $n$. If $\phi$ is further Hausdorff and second countable topological space, it is an $n$-manifold.}

Consider a hypersurface $\phi^i \mydef \big\{ \bd{r}(\bd{s})\in \Rc^{n+m} ~|~ s^i=c \big\}$ {\color{black} of the $n$-dimensional surface $\phi$} for a fixed index $i$ and a constant $c$. Define $\bigtriangledown s^i \in \Rc^{n+m}$ as the basic normal vector to the hypersurface $\phi^i$ {\color{black} in $T \phi_p$}, given  by\footnote{{\color{black}To have a  proper inverse map for $\bd{r}(\bd{s})$ when $m > 0$, one must restrict the inversion to the  
embedded submanifold $\bd{r}(\Rc^{n})$ contained in $\Rc^{n+m}$.}}
\begin{equation}
	\bigtriangledown s^i \mydef \bigg(\frac{\partial s^i}{\partial r^1}, \dots, \frac{\partial s^i}{\partial r^{n+m}}\bigg). \label{eq:basic_normal}
\end{equation}%
Then, $\{ \bigtriangledown s^i \}$ forms the dual basis of tangent space $T\phi_p$.

Hence, we have
\begin{equation}
	\langle \bd{r}_i, \bigtriangledown s^j \rangle = \delta_i^j, \label{eq:rbys}
\end{equation}%
where $\delta_i^j$ is the Kronecker delta.

\subsection{Metric Tensors}
The covariant metric tensor $(g_{ij})$ of the surface $\phi$ is a 2-dimensional symmetric matrix whose entries $g_{ij}$ are given by the inner products of the tangent vectors,
\begin{equation}
	g_{ij} \mydef \langle \bd{r}_i, \bd{r}_j \rangle. \label{eq:covariant}
\end{equation}%

Similarly, define the contravariant metric tensor $(g^{ij})$ of the surface $\phi$ by the normal vectors,
\begin{equation}
	g^{ij} \mydef \langle \bigtriangledown s^i, \bigtriangledown s^j \rangle. \label{eq:contravariant}
\end{equation}%

By \eqref{eq:rbys} we have,
\begin{equation}
	\sum_{k=1}^{n} g_{ik} g^{kj} = \delta_i^j,
\end{equation}%
which indicates that the contravariant metric tensor is the inverse matrix of the covariant metric tensor, namely, 
\begin{equation}
	\big( g^{ij} \big) = ( g_{ij} )^{-1}. \label{eq:identity}
\end{equation}%
We will implement \eqref{eq:identity} when evaluating the voltage stability boundary in Section~\ref{sec:proximity}. 

\subsection{Levi-Civita Connection}\label{subsec:connection}
The metrics discussed above are naturally Riemannian metrics. Equipped with these metrics, we can further discuss ``straight lines'' on the smooth manifold $\phi$. In order to do so, we need an appropriate tool to preserve the Riemannian metric, which, in Riemannian geometry, is the Levi-Civita connection. 

The coefficients of the Levi-Civita connection are locally characterized by the Christoffel symbols of the second kind, denoted by $\varGamma_{ij}^k$. In what follows, we shall first compute the Christoffel symbols of the first kind $\varGamma_{ij,k}$, 
\begin{subequations}
	\begin{align}
	\varGamma_{ij,k} &= \frac{1}{2} \bigg( \frac{\partial g_{jk}}{\partial s^i} + \frac{\partial g_{ik}}{\partial s^j} - \frac{\partial g_{ij}}{\partial s^k} \bigg) \label{eq:ch1_1}\\
	&= \langle \bd{r}_{ij},\bd{r}_k \rangle, \label{eq:ch1_2}
	\end{align}\label{eq:ch1}%
\end{subequations}%
where $\bd{r_{ij}} \mydef \partial \bd{r}_i/ \partial s^j$. Then, the second kind $\varGamma_{ij}^k$ is derived as,
\begin{subequations}
	\begin{align}
		\varGamma_{ij}^k &= \frac{1}{2} \sum_{l=1}^{n} g^{kl} \bigg( \frac{\partial g_{jl}}{\partial s^i} + \frac{\partial g_{il}}{\partial s^j} - \frac{\partial g_{ij}}{\partial s^l} \bigg) \label{eq:ch2_1}\\
		&=  \sum_{l=1}^{n} g^{kl} \varGamma_{ij,l}. \label{eq:ch2_2}
	\end{align}\label{eq:ch2}
\end{subequations}%

\begin{figure}[tb!]
	\begin{center}
		\subfigure[Less curved surface]{\label{fig:gamma_flat}\includegraphics[width=0.48\columnwidth]{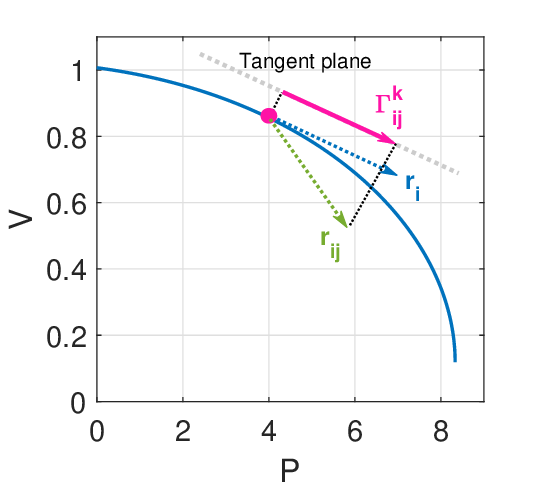}}
		\subfigure[More curved surface]{\label{fig:gamma_curve}\includegraphics[width=0.48\columnwidth]{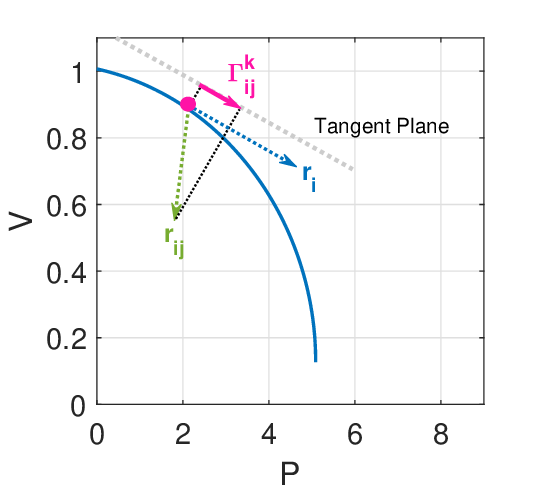}}
  	\end{center}
		\caption{Illustration of the Levi-Civita Connection.}
		\label{fig:gamma}
\end{figure}

{\color{black}
Geometrically speaking, if we project the rate of change of tangent vectors, $\bd{r}_{ij}$, to the tangent space $T \phi_p$, the representation under the basis of the tangent vectors are the coefficients of the Levi-Civita connection. 
\begin{equation}
	\mathscr{P}(\bd{r}_{ij}) = \sum_{k=1}^{n} \varGamma_{ij}^k \bd{r}_k
\end{equation}%
where $\mathscr{P}$ is the projection operator.

That is, $\varGamma_{ij}^k$ is the $k$-th projected component of the second order derivative $\bd{r}_{ij}$ onto the tangent space. For a constant $|\bd{r}_{ij}|$, if $\varGamma_{ij}^k$ is small, it means the component of $\bd{r}_{ij}$ along the normal direction is large, and this suggests a more curved surface.
For example, Fig.~\ref{fig:gamma_flat} shows a less curved surface (the blue curve) with a large $\varGamma_{ij}^k$ value. On the contrary, Fig.~\ref{fig:gamma_curve} shows a more curved surface with a small $\varGamma_{ij}^k$ value. }

\subsection{Geodesic Equation}
Consider a smooth curve $\gamma$ on the manifold $\phi$. It can be described by the canonical parametric representation in the local coordinate system $\bd{r}\big( \bd{s}(\tau) \big): \Rc \to \Rc^{n+m}$. For a curve to be a geodesic curve, that is, the shortest path between two points on a Riemannian manifold, $\gamma$ must satisfy the following geodesic equation that represents zero acceleration,
\begin{equation}
	\frac{\partial^2 s^k}{\partial \tau^2} + \sum_{i,j}^{n} \varGamma_{ij}^k \frac{\partial s^i}{\partial \tau} \frac{\partial s^j}{\partial \tau} = 0,  \ \forall k, \label{eq:geodesic}
\end{equation}%
where $\tau$ is a canonical parameterization which is proportional to the arc length of $\gamma$. Equation \eqref{eq:geodesic} serves as the core formula for us to approximate the voltage stability boundary in Section~\ref{sec:proximity}. 

\section{Approximate Voltage Stability Boundary}
   \label{sec:proximity}
In this section, we propose a novel method to approximate the voltage stability boundary based on geometric quantities discussed in Section~\ref{sec:geodesic}.  

\subsection{Approximating Voltage Stability Boundary}
\label{subsec:apprx_volt}
The interpretation in Section~\ref{subsec:interpretation} implies that by examining certain geometric properties of the chart $U_{\text{max}}$ we might be able to predict its boundary. Among different geometric objects, we pay special attention to the geodesic curve because it inherits the intrinsic geometry of the manifold and can be naturally expanded to the {\color{black}neighborhood of the operating point and the proximity of the singular boundary}. 

First, consider any curve in the chart $U_{\text{max}}$ starting from the origin $\bd{z}_0$ of the local coordinate system. Given the canonical parameterization, we can expand the voltage magnitude variable $V^k$ by the Taylor series,
\begin{equation}
    V^k(\tau) = V^k(0) + \sum_{n=1}^{\infty} \frac{\tau^n}{n!}\frac{d^n V^k(0)}{d \tau^n}, \label{eq:Taylor}
\end{equation}
where $V^k(0)$ is at the origin $\bd{z}_0$, $\tau^n$ denotes $\tau$ to the $n$-th power.

Let's further assume that the curve is geodesic. The higher order derivatives $d^n V^k(0)/d \tau^n$ in \eqref{eq:Taylor} can be evaluated by taking higher order derivatives of the Christoffel symbols from~\eqref{eq:geodesic}, 
\begin{align}
            V^k(\tau) & = V^k(0) + \dot{V}^k(0) \tau - \nonumber\\ &\sum_{n=2}^{\infty} \frac{\tau^n}{n!} \sum_{i_1,\cdots,i_n}^{N} \varGamma_{i_1 \cdots i_n}^k \dot{X}^{i_1}(0) \cdots \dot{X}^{i_n}(0),  \label{eq:geod_Taylor}%
\end{align}%
where $\bd{X}$ includes both voltage magnitude $\bd{V}$ and angle $\bd{\ttt}$, $X^i$ is the $i$-th entry of $\bd{X}$, $\dot{X}^i$ is the derivative of $X^i$ with respect to $\tau$, and $\varGamma_{i_1 \cdots i_n}^k$ represents generalized connections \cite{brewin2009:riemann}.
 
To simplify the model and improve computational efficiency, we truncate \eqref{eq:geod_Taylor} to the second order, namely,
\begin{equation}
    V^k(\tau) \approx V^k(0) + \dot{V}^k(0) \tau - \frac{ \tau^2 }{2} \sum_{i,j}^{N} \varGamma_{ij}^k \dot{X}^i(0) \dot{X}^j(0) .\label{eq:geod_Taylor_2nd}
\end{equation}%

As an implication of Taylor's theorem, \eqref{eq:geod_Taylor_2nd} closely matches the behavior of the geodesic curve in the vicinity of the origin. We extend the approximation for estimating the boundary of the chart. In particular, \eqref{eq:geod_Taylor_2nd} admits a unique $\tau^\star$ (that is independent of the origin $V^k(0)$) which yields the extremum of the univariate linear and quadratic parts of \eqref{eq:geod_Taylor_2nd}. Therefore, by substituting $\tau^\star$ into \eqref{eq:geod_Taylor_2nd} and enforcing the sign of the slope $\dot{V}^k(0)$ on the first and second order terms, we provide the approximated voltage stability boundary $V^k_{appx}$ for the $k$-th PQ bus as follows,
\begin{equation}
    V^k_{appx} = V^k(0) +  \frac{\dot{V}^k(0)^3}{|\dot{V}^k(0)|} \bigg( 2\sum_{i,j}^{N} \varGamma_{ij}^k \dot{X}^i(0) \dot{X}^j(0) \bigg)^{-1}.\label{eq:volt_boundary}%
\end{equation}%

Extensive numerical simulations in Section~\ref{sec:num} will show that the approximated boundary from \eqref{eq:volt_boundary} can follow the shape of the true boundary consistently and conservatively. Considering this approximation is constructed based on a single operating point, its ability to estimate the shape of the boundary in global directions is impressive. {\color{black} Although rigorous proof of this result is under investigation, we provide a geometric interpretation for this phenomenon here.} 
Recall Fig.~\ref{fig:gamma} and Section~\ref{subsec:connection}, a smaller Christoffel symbol $\varGamma_{ij}^k$ implies that the manifold is more curved in direction-$k$. If the geodesic path is strongly inclined to direction-$k$, it should also be more curved in this direction. So, the voltage magnitude $V^k$ at bus-$k$ should experience a deeper drop on the stability boundary. Meanwhile, with a smaller $\varGamma_{ij}^k$ the quadratic approximation \eqref{eq:geod_Taylor_2nd} has a smaller coefficient for the quadratic term. Hence, \eqref{eq:volt_boundary} predicts a deeper voltage decline at bus-$k$. 
By this means, \eqref{eq:volt_boundary} can consistently follow the shape of the true boundary.

In summary, the general procedure of the proposed method is given as follows:
\begin{enumerate}
	\item Generate the covariant metric tensor $(g_{ij})$ from the first-order derivatives using \eqref{eq:covariant}.
	\item Obtain the contravariant metric tensor $(g^{ij})$ by taking the inverse matrix of $(g_{ij})$ using \eqref{eq:identity}.
	\item Compute the Christoffel symbols of the first kind $\varGamma_{ij}^k$ using \eqref{eq:ch1_2}.
	\item Compute the Christoffel symbols of the second kind using \eqref{eq:ch2_2}.
        \item Approximate voltage stability boundary on each bus using \eqref{eq:volt_boundary}.
\end{enumerate}

\subsection{Improving Conservativeness} \label{subsec:conserv}
{\color{black}
We further provide a heuristic to improve the performance of the above estimation. This heuristic is motivated by the consistent approximation gap (c.f. Section~\ref{sec:num}) that is nearly uniform on the same bus for all power-varying directions. This insightful observation motivates the following modification of our method.
Since \eqref{eq:volt_boundary} consistently yields a conservative estimation of the voltage stability boundary,  
instead of working on \eqref{eq:volt_boundary}, let us consider,
\begin{equation}
	V^k_{appx} = V^k(0) +  \alpha^k \frac{ \dot{V}^k(0)^3}{|\dot{V}^k(0)|} \bigg( 2\sum_{i,j}^{N} \varGamma_{ij}^k \dot{X}^i(0) \dot{X}^j(0) \bigg)^{-1}, \label{eq:volt_boundary_mod}%
\end{equation}%
where $\alpha^k \in \Rc$ is a scaling factor that can be estimated by, for example, running a continuation for a specific power-varying direction. When $\alpha^k=1$, \eqref{eq:volt_boundary_mod} reverts back to the original form in \eqref{eq:volt_boundary}. This modification will substantially improve the conservativeness of the original approximation which shall be seen in the second part of Section~\ref{sec:num}.}

\section{Numerical Simulations}
	\label{sec:num}
This section presents extensive numerical results to demonstrate the performance of the proposed method under various settings and scenarios. We implement continuation power flow (CPF) as the benchmark method to identify a collection of nose points as the true voltage stability boundary. 

{\color{black}
In the first subsection, we highlight the capability of the original model \eqref{eq:volt_boundary} to reveal the global geometry (in a conservative manner) of the voltage stability boundary based on a single operating point. The second subsection further illustrates how the modified approach \eqref{eq:volt_boundary_mod} improves the conservativeness and accuracy of the voltage stability boundary estimation. In the third subsection, we highlight the computational efficiency of the proposed method. All the simulations are performed using Matlab 2017b on a PC with a 4-core $2.8$ GHz CPU. 
}

In what follows, blue curves represent the true voltage stability boundary obtained from the CPF, green curves are the approximated voltage stability boundary from the proposed methods, and black lines are the starting points.  

{\color{black} 
\subsection{Simulation Results of the Original Method}
This subsection demonstrates the capability of the original model \eqref{eq:volt_boundary} to capture the global shape of the voltage stability boundary using both 2D and 3D visualizations of two benchmark systems, i.e., the IEEE 14-Bus and 39-Bus systems. Note that the proposed methods are suitable for any dimensional problems, but visualization can only be realized in at most 3 dimensions. 

To mimic the high variability of renewable generation, we superimpose faster renewable variations on top of slower load changes in all possible directions. The time-varying power injection model is given by
\begin{subequations}
	\begin{align}
		P^i(t) &= P^i_0 + \cos(\beta) t, \\
		P^j(t) &= P^j_0 + \sin(\beta) t,
	\end{align} \label{eq:direction}%
\end{subequations}%
where $P^{i}_0$ and $P^{j}_0$ are the initial power injections at node $i$ and $j$, respectively. Parameter $\beta$ is distributed evenly in the range $[0, 2 \pi]$, representing different power-varying directions. We simulate load change and renewable variation in the same way. When considering three different changing directions, two independent angle variables $\beta$ and $\delta$ are needed. 
\begin{subequations}
	\begin{align}
		P^i(t) &= P^i_0 + \cos(\beta) \cos(\delta) t, \\
		P^j(t) &= P^j_0 + \sin(\beta) \cos(\delta)  t, \\
		P^{k}(t) &= P^{k}_0 + \sin(\delta)  t.
	\end{align} \label{eq:direction_3d}%
\end{subequations}%

For the IEEE 14-Bus example, we select Bus-4 and Bus-9 as the load-varying nodes with a constant power factor at $0.95$ and select Bus-3 and Bus-6 as the renewable fluctuating nodes with a $4 \times$ faster-changing rate than the load changing rate. 

Figure~\ref{fig:case14_buses} shows where the voltage magnitude reaches its stability boundary at different buses. From a given operating point, the original model can replicate the global shape of the voltage stability boundary at each bus in a conservative and uniform manner. The approximated boundaries (green curves) follow the true boundaries (blue curves) consistently. 

Figure~\ref{fig:case14_ini_pt} presents the approximated boundaries using the original model at different initial points (i.e., different loading conditions). In the light loading condition (c.f. Figure~\ref{fig:case14_bus4_3}), the starting voltage magnitude at Bus-4, as shown by the black curve, is around $0.95$ p.u. In the heavy loading condition, (c.f. Figure~\ref{fig:case14_bus4_4}), the starting voltage magnitude at Bus-4 is below $0.9$ p.u. In both loading conditions, the approximated boundaries follow the true boundary consistently.  

To further illustrate the capability of the proposed method in a higher dimensional space, we add one more load-varying node at Bus-$11$ and one more renewable varying node at Bus-$2$. Figure~\ref{fig:case14_3d} shows the corresponding voltage stability boundary in 3D. One can observe that the high-dimensional approximation still captures the essential shape of the true voltage stability boundary. 

\begin{figure}[tb!]
	\begin{center}
		\subfigure[Voltage magnitude at Bus-4]{\label{fig:case14_bus4}\includegraphics[width=0.48\columnwidth]{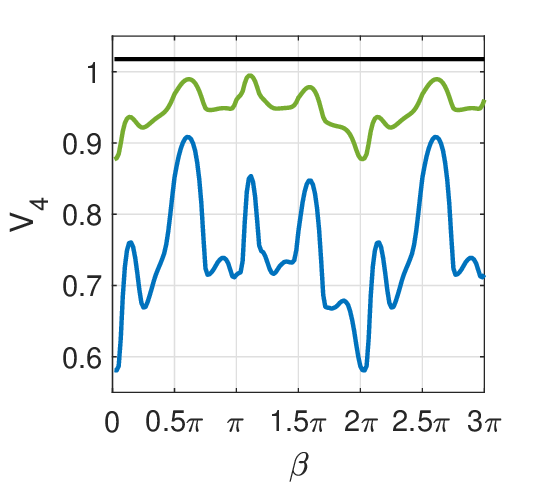}}
		\subfigure[Voltage magnitude at Bus-9]{\label{fig:case14_bus9}\includegraphics[width=0.48\columnwidth]{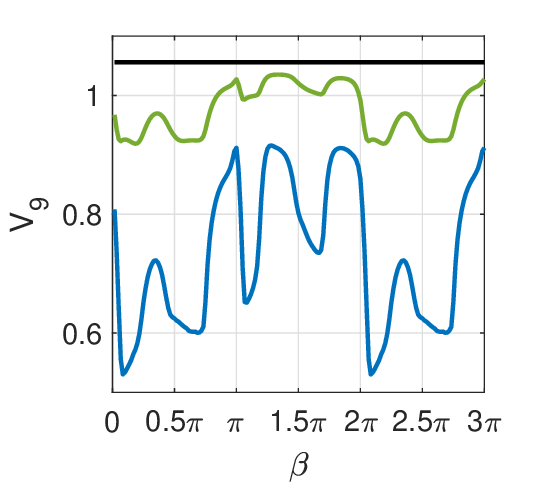}}
  	\end{center}
		\caption{IEEE 14-bus voltage stability boundary at different buses.}
		\label{fig:case14_buses}
\end{figure}

\begin{figure}[tb!]
	\begin{center}
		\subfigure[Light loading condition]{\label{fig:case14_bus4_3}\includegraphics[width=0.48\columnwidth]{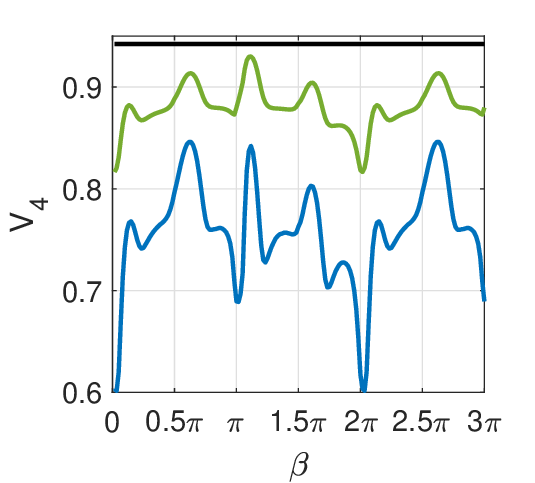}}
		\subfigure[Heavy loading condition]{\label{fig:case14_bus4_4}\includegraphics[width=0.48\columnwidth]{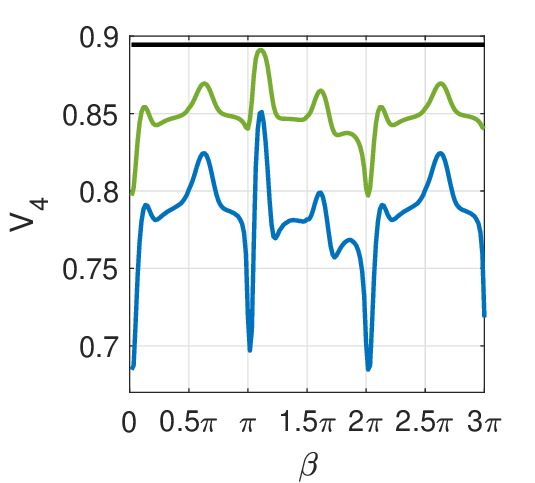}}
  	\end{center}
		\caption{IEEE 14-bus voltage stability boundary in different conditions.}
		\label{fig:case14_ini_pt}
\end{figure}

\begin{figure}[tb!]
	\begin{center}
		\subfigure[Authentic shape]{\label{fig:case14_3d_1}\includegraphics[width=0.48\columnwidth]{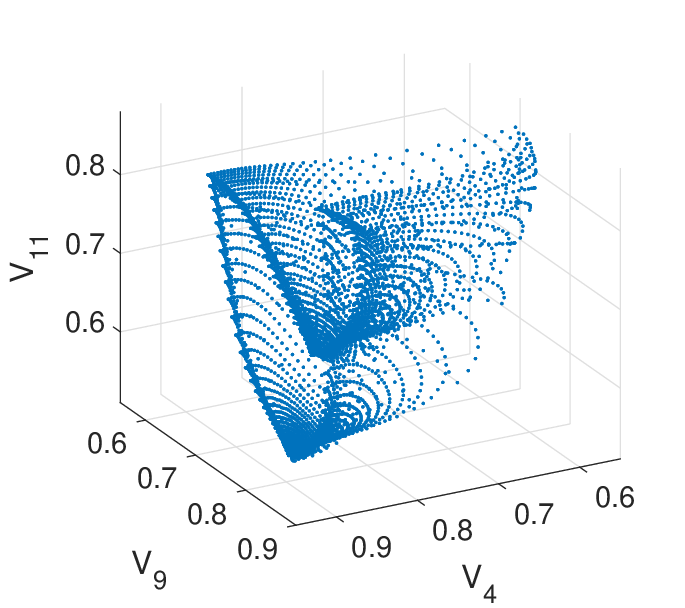}}
		\subfigure[Approximated shape]{\label{fig:case14_3d_2}\includegraphics[width=0.48\columnwidth]{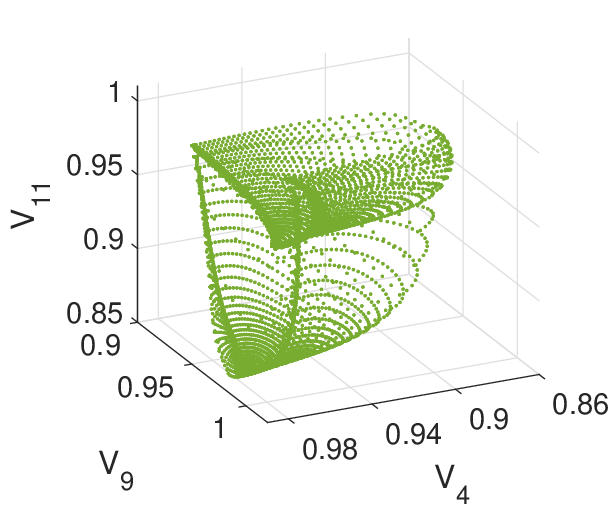}}
		\caption{IEEE 14-bus voltage stability boundary in 3D.}
		\label{fig:case14_3d}
	\end{center}
\end{figure}

For the IEEE 39-Bus example, we select Bus-4 and Bus-8 as the load-varying nodes with a constant power factor at $0.95$ and select Bus-33 and Bus-36 as the renewable fluctuating nodes with a $4 \times$ faster-changing rate than the load changing rate. 

Figure~\ref{fig:case39_buses} shows the approximated boundaries of the IEEE-39 example at Bus-4 and Bus-8, from a given operating point. Figure~\ref{fig:case39_ini_pt} depicts the approximated boundaries at different initial loading conditions. We further add Bus-21 as the third load-varying bus and plot the corresponding 3D voltage stability boundaries in Fig.~\ref{fig:case39_3d}. 

All the results obtained for the IEEE 39-Bus system share the same properties as those of the IEEE 14-Bus system. It suggests that the proposed model \eqref{eq:volt_boundary} can capture the global shape of the true voltage stability boundary in a consistent and conservative manner.

\begin{figure}[tb!]
	\begin{center}
		\subfigure[Voltage magnitude at Bus-4]{\label{fig:case39_bus4}\includegraphics[width=0.48\columnwidth]{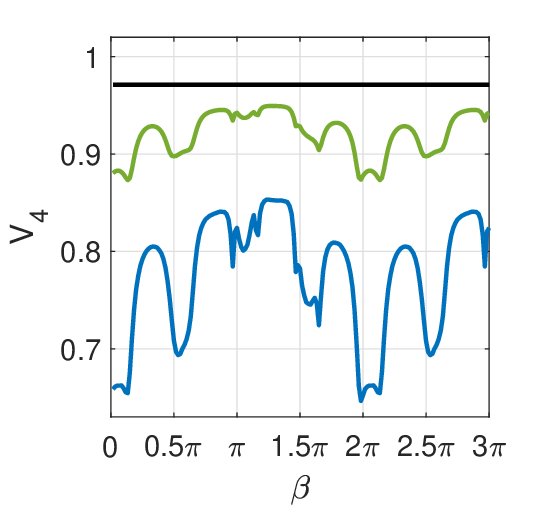}}
		\subfigure[Voltage magnitude at Bus-8]{\label{fig:case39_bus8}\includegraphics[width=0.48\columnwidth]{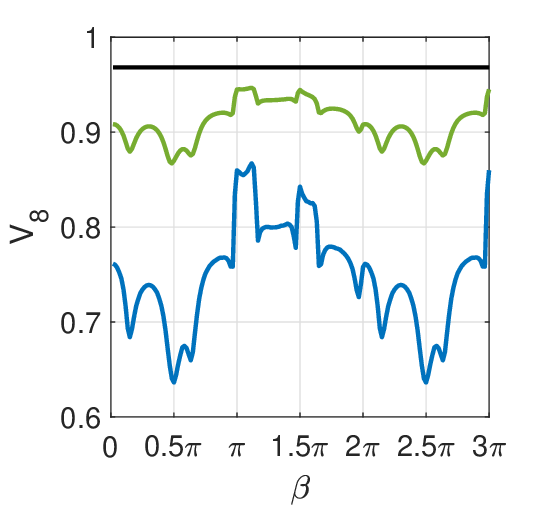}}
  	\end{center}
		\caption{IEEE 39-bus voltage stability boundary at different buses.} 
		\label{fig:case39_buses}
\end{figure}

\begin{figure}[tb!]
	\begin{center}
		\subfigure[Light loading condition]{\label{fig:case39_bus4_1}\includegraphics[width=0.48\columnwidth]{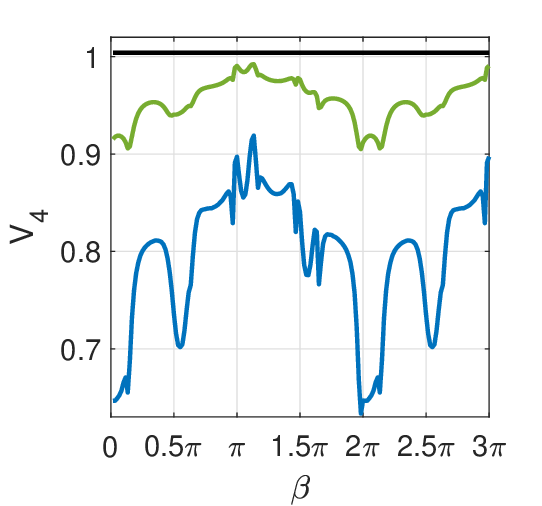}}
		\subfigure[Heavy loading condition]{\label{fig:case39_bus4_4}\includegraphics[width=0.48\columnwidth]{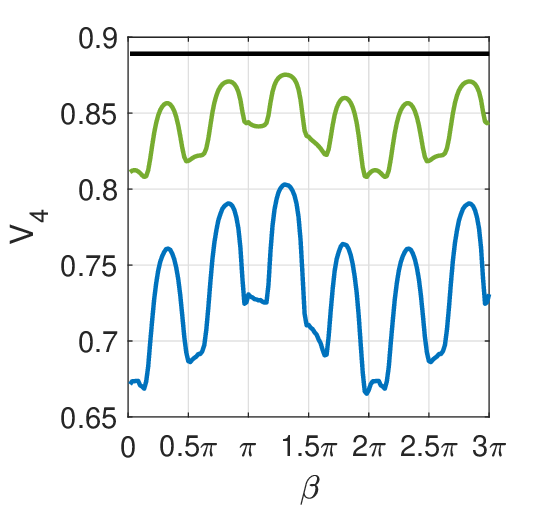}}
  	\end{center}
		\caption{IEEE 39-bus voltage stability boundary in different conditions.}
		\label{fig:case39_ini_pt}
\end{figure}

\begin{figure}[tb!]
	\begin{center}
		\subfigure[Authentic shape]{\label{fig:case39_3d_1}\includegraphics[width=0.48\columnwidth]{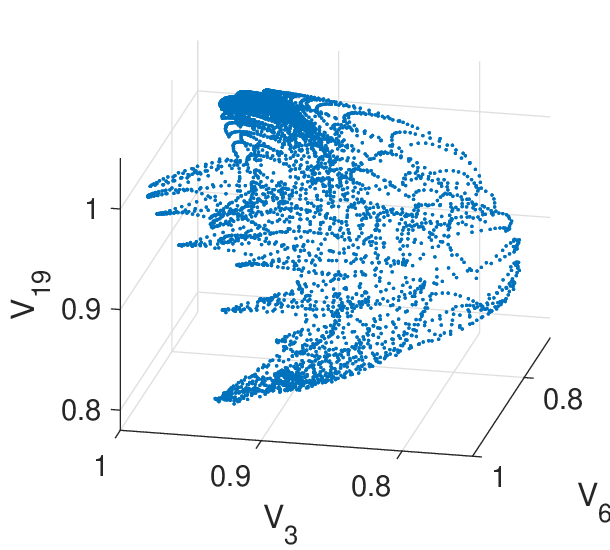}}
		\subfigure[Approximated shape]{\label{fig:case39_3d_2}\includegraphics[width=0.48\columnwidth]{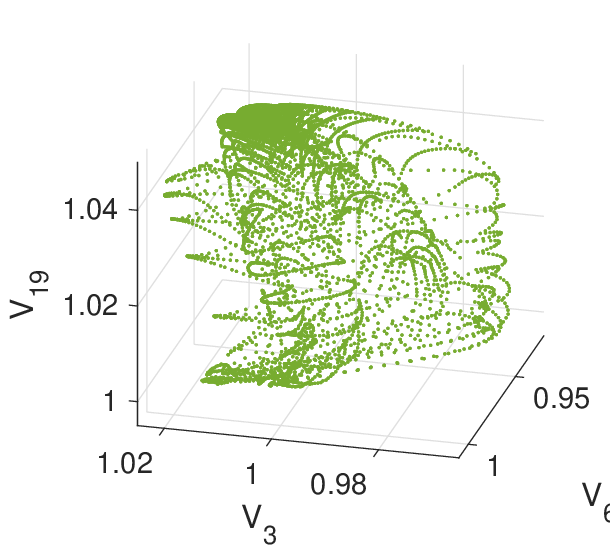}}
  	\end{center}
		\caption{IEEE 39-bus voltage stability boundary in 3D.}
		\label{fig:case39_3d}
\end{figure}

\subsection{Simulation Results of the Modified Method}
This subsection presents the simulation results from the modified model \eqref{eq:volt_boundary_mod} with the scaling factor $\alpha^k$ for Bus-$k$ to demonstrate the improved performance. 

Figures~\ref{fig:case14_diff_scale} and \ref{fig:case39_diff_scale} compare the modified approximation results with $\alpha^k \neq 1$ (the right column) to the original approximation (the left column) for the 14-Bus and 39-Bus systems, respectively. The plots demonstrate that a uniform scaling factor $\alpha^k$ can greatly improve the conservativeness of the original approximation for all power-varying directions $\beta$ at the same bus. In practice, to determine the value of this scaling factor $\alpha^k$, one can compute a single voltage collapse point through the CPF method to find a specific $\alpha^k$. Then, use this $\alpha^k$ for other power-varying directions. 

\begin{figure}[tb!]
	\begin{center}
		  \subfigure[Bus-4 voltage at $\alpha^4=1$]{\label{fig:case14_bus4_1x}\includegraphics[width=0.48\columnwidth]{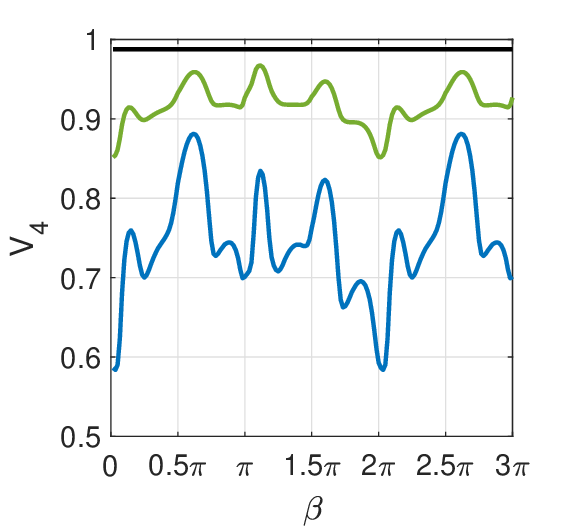}}
		  \subfigure[Bus-4 voltage at $\alpha^4=3.4$]{\label{fig:case14_bus4_34x}\includegraphics[width=0.48\columnwidth]{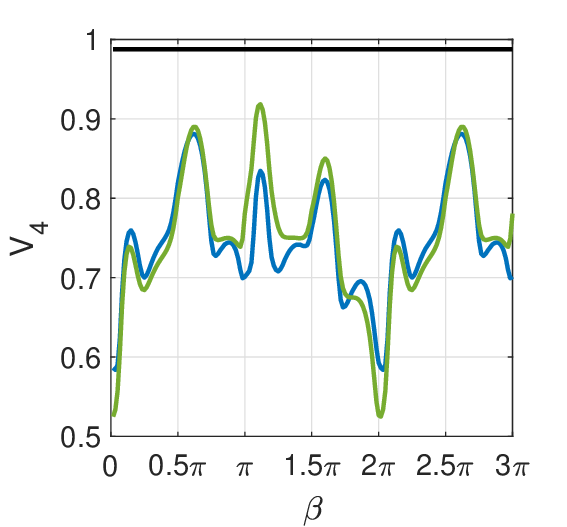}}\\
		  \subfigure[Bus-5 voltage at $\alpha^5=1$]{\label{fig:case14_bus5_1x}\includegraphics[width=0.48\columnwidth]{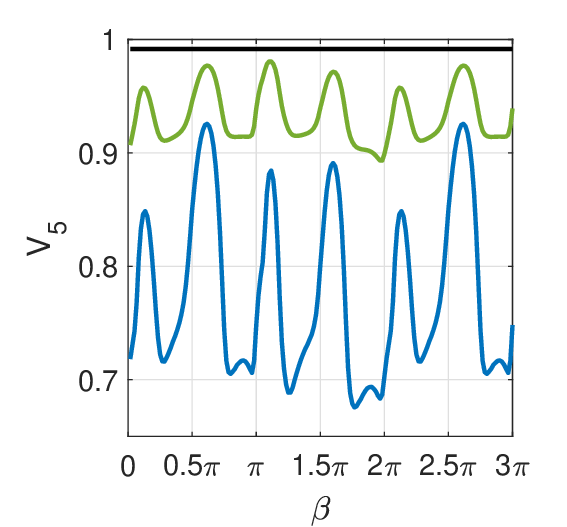}}
		  \subfigure[Bus-5 voltage at $\alpha^5=3.4$]{\label{fig:case14_bus5_34x}\includegraphics[width=0.48\columnwidth]{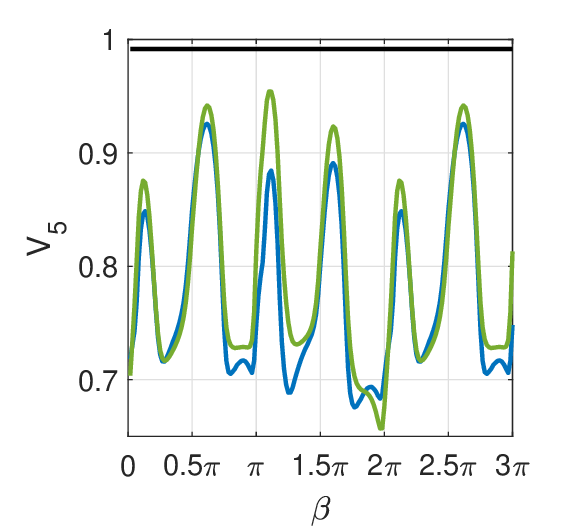}}
		\end{center}
        \caption{IEEE 14-Bus system with different scaling factors $\alpha^k$.}
		\label{fig:case14_diff_scale}
\end{figure}

\begin{figure}[tb!]
	\begin{center}
		  \subfigure[Bus-4 voltage at $\alpha^4=1$]{\label{fig:case39_bus4_1x}\includegraphics[width=0.48\columnwidth]{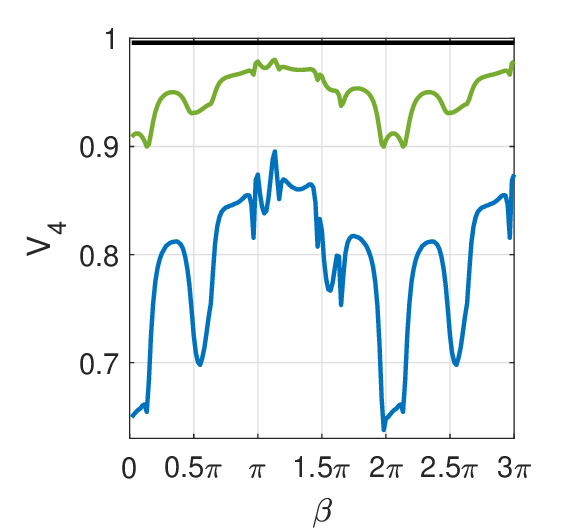}}
		  \subfigure[Bus-4 voltage at $\alpha^4=3.4$]{\label{fig:case39_bus4_34x}\includegraphics[width=0.48\columnwidth]{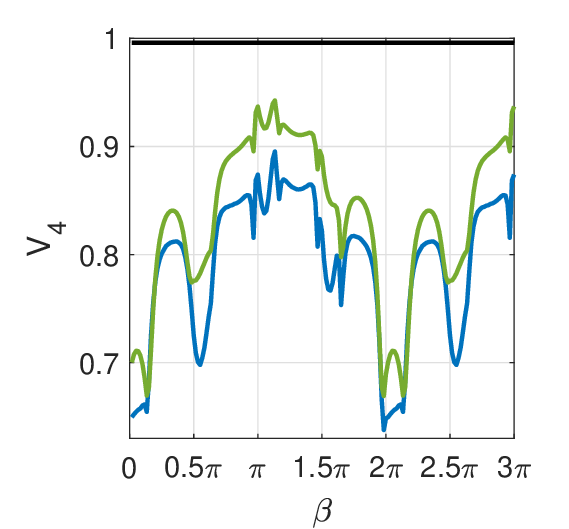}}\\
		  \subfigure[Bus-8 voltage at $\alpha^8=1$]{\label{fig:case39_bus8_1x}\includegraphics[width=0.48\columnwidth]{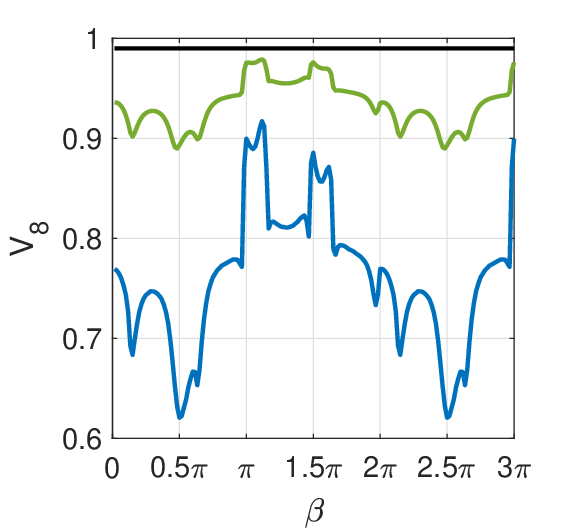}}
		  \subfigure[Bus-8 voltage at $\alpha^8=3.4$]{\label{fig:case39_bus8_34x}\includegraphics[width=0.48\columnwidth]{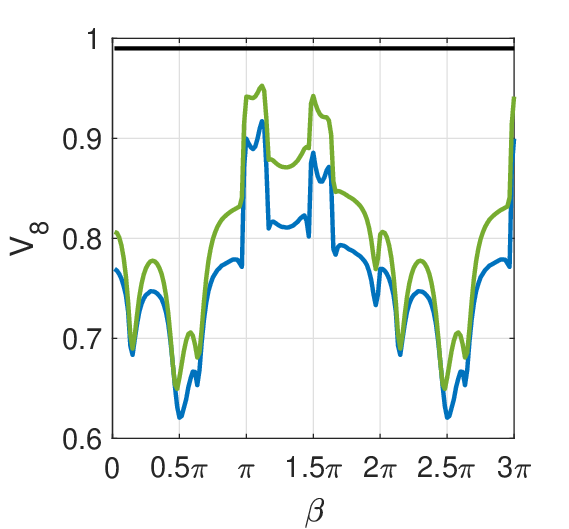}}
        \end{center}
		\caption{IEEE 39-Bus system with different scaling factors $\alpha^k$.}
		\label{fig:case39_diff_scale}
\end{figure}
}

\subsection{Computational Efficiency}
We summarize the execution time for different IEEE benchmark systems in Table~\ref{tab:time}. Each system is evaluated under $180$ different power-varying directions on a circle for a fixed initial condition. The speedup of the proposed method in our implementation is around $1000 \times$ compared to the CPF method. It suggests that the proposed method can be used for online voltage stability boundary analysis under high variability of renewable. Further efficiency improvement is promised when the tensor sparsity is explored. 
\begin{table}[tb!]
	\centering
	\caption{Execution time for IEEE benchmark systems.}
	\begin{tabular}{c|c|c|c}
		\hline \hline
		Bus     & 9-Bus & 14-Bus & 39-Bus  \\ \hline
		CPF (sec)    & 10.4   & 17.5   & 42.5    \\ \hline
		Proposed (sec) & 0.0141 & 0.0191 & 0.0441  \\ \hline
		Speedup & 739    & 916    & 964       \\ \hline \hline
	\end{tabular} \label{tab:time}%
\end{table}%
	
\section{Discussions}
	\label{sec:discussion}
One of the most intriguing results is that the proposed method can reconstruct the basic geometry of the voltage stability boundary from a single operating point. Geometrically speaking, the Christoffel symbols are evaluated at a given point. Then, the geodesic equation \eqref{eq:geodesic} extends the local information to the neighborhood of the given point (which is still in the local sense). However, the collection of the extrema obtained from the truncated quadratic approximation \eqref{eq:volt_boundary} shows its capability of replicating the global shape of the true voltage stability boundary in a conservative way. 
Understanding such resemblance may inspire completely new mathematical tools for analyzing global behaviors of future complex power grids based on local measurements and computations. 

\section{Conclusion} 
   	\label{sec:concl}
This paper discusses how to use local geometry to approximate the shape of voltage stability boundary for future renewable-rich power grids with large spatial-temporal disturbances. Instead of evaluating voltage stability in 1D predefined directions, we first extracted intrinsic geometric information, i.e., the Christoffel symbols for the Levi-Civita connection, of the power flow solution manifold at a given operating point. 
Then, the geodesic equation was further evaluated to approximate the voltage magnitude on each bus in the univariate quadratic form. The extrema of these univariate quadratic equations were used to approximate the voltage stability boundary and a further modification was proposed to substantially improve the conservativeness. Extensive numerical simulations under different scenarios were carried out on IEEE 14-Bus and 39-Bus systems to show the accuracy of the proposed method in replicating the true shape of the stability boundary in the global sense. 
It also showed a $1000 \times$ speedup of computational efficiency compared to the continuation power flow method. Thus, the proposed method is particularly suitable to handle high-dimensional wide-range variability in renewable-rich power grids.

One of the future research directions will be focused on exploring the sparsity of the proposed method to further enhance the computational efficiency. Another interesting topic is to show why the global geometry of the voltage stability boundary is embedded in the local Levi-Civita connections. 

\bibliographystyle{IEEEtran}
\bibliography{RefVS}


 




\vfill

\end{document}